\documentclass[12pt,fleqne]{article}
\usepackage{latexsym,amsfonts}

\usepackage{graphicx}

\newcommand{\be} {\begin{equation}}
\newcommand{\ee} {\end{equation}}
\newcommand{\ba} {\begin{array}}
\newcommand{\ea} {\end{array}}
\newcommand{\p} {\partial}

\newcommand{\vp} {\varphi}

\newcommand{\var} {\varepsilon}
\newcommand{\al} {\alpha}

\newcommand{\lbd} {\lambda}
\newcommand{\ds} {\displaystyle}
\newcommand{\lo} {\left (}
\newcommand{\ro} {\right)}
\newcommand{\ve} {\varepsilon}

\begin{document}

\title{\textbf{Solitary wave and other solutions\\ for nonlinear heat
equations}}
\author{\textbf{Anatoly G. Nikitin and Tetyana A. Barannyk} \\
Institute of Mathematics, National Academy of
Sciences\\
of the Ukraine, 3 Tereshchenkivska Street, Kiev 4,
Ukraine} \maketitle
\abstract{New exact solutions for the heat equation with a
polynomial non-linearity and for the Fisher equation are
found. An extended class of non-linear heat equations admitting solitary wave
solutions
is found. The generalization of the Fisher equation is proposed
whose solutions propagate with arbitrary ad hoc fixed velocity.}

\renewcommand{\theequation}{\arabic{section}.\arabic{equation}}
\setcounter{equation}{0}

\section{Introduction}

The nonlinear reaction-diffusion equations play fundamental role in
a great number
of various models of heat and reaction-diffusion processes,
mathematical biology, chemistry, genetics and many, many others.
 Thus, one of the corner stones of mathematical biology is
the Fisher equation \cite{Murry}
\be \label{1.1}
u_t-u_{xx}=u(1-u)
\ee
where $u=u(x,t)$ and subscripts denote derivatives w.r.t. the
corresponding variable: $u_t=\frac{\p u}{\p t}$,
$u_{xx}=\frac{\p^2u}{\p x^2}$.

Equation (\ref{1.1}) is a particular case of the
Kolmogorov-Petrovskii-Piskunov (KPP) equation \cite{KPP}
\be\label{KPP}
u_t-u_{xx}=f(u) \ee where $f(u)$ is a sufficiently smooth function
satisfying the relations $f(0)=f(1)=0,\ f_u(0)=\al>0,\
f_u(u)<\al,\ 0<u<1$.

The reaction-diffusion equation with the cubic polynomial
nonlinearity \be \label{1.3} u_t-u_{xx}=\al(u^3+bu^2+cu) \ee
where $\al=\pm 1$, $b$ and $c$ are constants, also has a large
application value and includes as particular cases the
Fitzhugh-Nagumo equation \cite{Fit}
 $(\al=-1, b=-c-1, 0<c<1)$ which is used in
population genetics, the Newell-Whitehead \cite{Newell}
 $(\mbox{for}\ c = \al=-1, b=0)$
and Huxley \cite{Murry} $(\mbox{for} \ \al=b=-1, c=0)$ equations.
Notice that the Fitzhugh-Nagumo equation also belongs to the
Kolmogorov-Petrovskii-Piskunov type.

A nice property of equations (\ref{1.1})-(\ref{1.3}) is that they
admit plane wave solutions which in many cases can be found in
explicit form. Existence of such solutions is caused by the
symmetry w.r.t. translations $t\to t+k$,
$x \to x+ r$ with constant parameters $k$ and $r$. For some special
functions $f(u)$  equation admits more extended symmetry
groups \cite{Dorodnitsyn} and, as a result, have exact solutions
of more general type than plane waves. We notice that group
analysis of (\ref{KPP}) for $f(u)=0$ was carried out by Sophus Lie
more than 130 year ago \cite{Lie}. The group classification of
systems of nonlinear heat equations was presented in papers
 \cite{Ron}.

The conditional (non-classical) symmetry approach \cite{Bluman},
\cite{Fush}, \cite{Wint} enables to
construct new exact solutions of partial differential equations
which
cannot be found in the framework of Lie theory. In application to the
 equations of type (\ref{KPP}) this approach was successfully
used for the case of cubic polynomial nonlinearity
(\ref{1.3}) only. A systematic study of conditional symmetries
of equation (\ref{1.3}) was started by Fushchych and Serov
\cite{Fush2} whereas the most exhaustive analysis of these
symmetries was presented by Clarkson and Mansfield \cite{Clarkson}.
In particular, a number of exact solutions for (\ref{1.3}) was
found in \cite{Fush2}, \cite{Clarkson}.

An effective algorithm for construction of traveling wave solutions
together with a number of interesting examples
was proposed in the recent paper \cite{fan}. However the nonlinear
heat
equations of the
general type (\ref{KPP}) were not analyzed in \cite{fan}.

A goal of our paper is to add the list of known exact solutions of
equation (\ref{1.3}). Effectively we present an infinite number of
them. In addition, using the unified algebraic method \cite{fan}
we select such equations of the type (\ref{KPP})
which admit solitary wave solutions and construct these solutions
explicitly. Finally, we present solutions for the Fisher equation
and propose such generalization of it which admit
the same exact traveling wave solution as (1.1), but {\it with
any ad hoc given  velocity of propagation} (for solutions of (1.1)
this velocity is fixed and equal to $\frac{5}{\sqrt 6}$). In
spirit of Hirota's method \cite{Hirota} to achieve these goals we
use a special Ans\"atz which leads to a uniform formulation for
all considered equations (which, however, is {\it tri-linear}).
In
addition to equations  (\ref{1.1}) and (\ref{1.3}), this Ans\"atz
makes it possible to reduce an extended class of equations of the
type (\ref{KPP}).

In the following section we present a specific Ans\"atz which will
be used to reduce a class of nonlinear heat equations.In Section 3
we find an infinite set of new explicit elliptic
solutions for the heat equation with the cubic and cubic
polynomial nonlinearity.

In Section 4 we describe plane wave solutions for special classes
of equations (\ref{KPP}). In Section 5 solitary wave solutions for
equations (\ref{KPP}) are found. Finally, in Sections 6 and 7 we
present
 exact solutions for the Fisher equation and propose
a generalization of this equation.

\renewcommand{\theequation}{\arabic{section}.\arabic{equation}}
\setcounter{equation}{0}

\section{The Ans\"atz and related equations}

We start with the reaction-diffusion equation with a power
nonlinearity \be \label{2.1}  u_t-u_{xx}=- \lbd u^n, \quad \lbd
=\frac{2(n+1)}{(n-1)^2} \ee
where $n$ is a constant, $n\neq 1 $.

For convenience we choose a special value for the coupling
constant $\lbd$. Scaling  $u$ one can reduce
$\lbd$ to $1$ or to -1 for $n>1$ and $n<1$ respectively.

For any $n\not=1$ we set
\be \label{2.2}
u=\left(\frac{z_x}{z}\right)^k, \quad k={\frac{2}{n-1}}
\ee
and transform (\ref{2.1}) to the uniform equation
\be \label{2.3}
z\left(z_x z_{tx}-z_x z_{xxx} -(k-1)z^2_{xx}\right)=
z_x^2\left(z_t-(2k+1)z_{xx}\right).
\ee

In contrast with (\ref{2.1}) equation (\ref{2.3}) is homogeneous
with respect to the dependent variable and includes the cubic
non-linearities only while (\ref{2.1}) includes $u$ in
an arbitrary (fixed) power $n$. We will show that formulation
(\ref{2.3}) is very convenient for effective reductions.

We notice that (\ref{2.1}) is not the only nonlinear equation of
type (\ref{KPP}) which can be reduced to the tri-linear form via
Ans\"atz (\ref{2.2}). A more general equation (\ref{KPP}) which
admits this
 procedure is
\be \label{2.4} u_t-u_{xx}=k\left(-(k+1)u^n+ \lbd_1 u+ \lbd_2
u^\frac{n+1}{2}+
 \lbd_3 u^\frac{3-n}{2}+ \lbd_4 u^{2-n}\right)
\ee
where $\lbd_1, \ldots, \lbd_4$ are arbitrary constants.

Formula (\ref{2.4}) defines special but rather extended class of
the nonlinear heat equations, which includes all important models
enumerated in Introduction and many others. A specific presentation
of the coupling constants is chosen for convenience. The
change (\ref{2.2}) transforms (\ref{2.4}) to
the following form \be \label{2.6} \ba{l}
z(z_x\dot z_x-z_xz_{xxx}-\lbd_3 zz_x-\lbd_4 z^2-(k-1)z^2_{xx})\\
=z^2_x(\dot z+\lbd_1 z+\lbd_2 z_x-(2k+1)z_{xx}).
\ea
\ee

In contrast with (\ref{2.4})  equation (\ref{2.6}) is homogeneous
w.r.t. the dependent variable and is much more convenient for searching
for exact solutions.
\renewcommand{\theequation}{\arabic{section}.\arabic{equation}}
\setcounter{equation}{0}

\section{Infinite sets of solutions}

Consider a particular (but important) case of (\ref{2.4}) which
corresponds to
$n=3, \lbd_1=\lbd_2=\lbd_3=0$:
\be
\label{4.2} \dot u-u_{xx}=-2u^3. \ee
The related equation (\ref{2.6}) takes the form
 \be
\label{4.1} z(\dot z_x-z_{xxx})=z_x(\dot z-6z_{xx}). \ee

\begin{figure}[th]
\centerline{\includegraphics{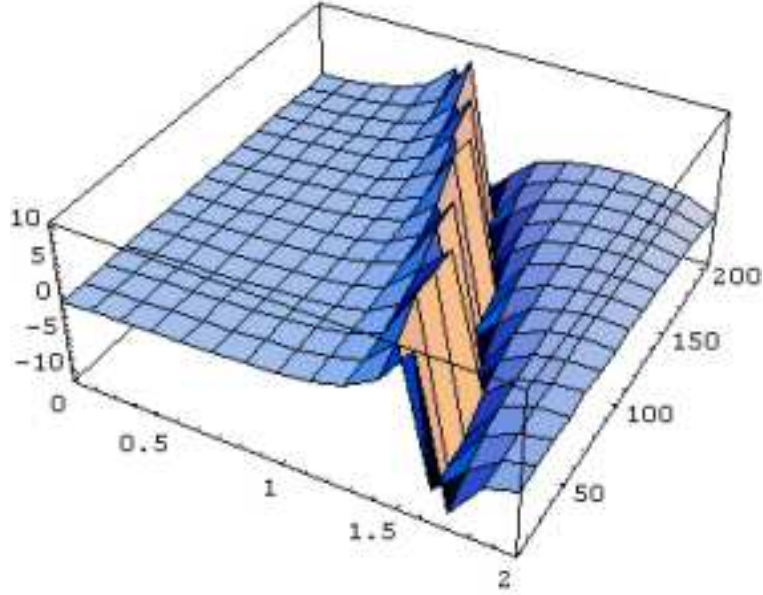}}
\caption{Solution (\ref{4.3})  for equation (\ref{4.2})} \label{f1}
\end{figure}

 Using the conditional symmetry approach the following exact
solution for
(\ref{4.2}) was found \cite{Clarkson} \be \label{4.3} u=2x {\rm
ds}\lo x^2+6t, \frac{1}{\sqrt 2}\ro \ee where ${\rm ds}(y,k)$ is
the Jacobi elliptic function satisfying \[ \left(
\frac{d\eta}{dy}\right)^2=k^2(k^2-1)+(2k^2-1)\eta^2+\eta^4. \]
The plot of this solution for $t<200$ created with MATHEMATICA is given in Fig. \ref{f1}.

Here we present other elliptic functions solutions for (\ref{4.2}),
effectively an infinite number of
them. To achieve this goal we exploit conditional symmetry of the
potential equation (\ref{4.1}).

Equation (\ref{4.1}) is compatible with the condition $Xz=0$ where
\be \label{Cond} X=\frac{\partial}{\partial t}- \frac {3}{x}
\frac{\partial}{\partial x}.\ee It means \cite{Fush} that this
equation admits conditional symmetry , thus it is reasonable to
search for its solutions in the form
\be \label{4.5} z=\vp(y),
\quad y=x^2+6t \ee
where $y$ is the invariant variable for
symmetry (\ref{Cond}). Substituting (\ref{4.5}) into (\ref{4.1})
we come to the third order differential equation for $\vp$ \be
\label{4.6} \vp \vp_{yyy}=3\vp_y \vp_{yy}. \ee

Dividing the l.h.s. and r.h.s. of (\ref{4.6}) by $\vp \vp_{yy}$ and
integrating we obtain
\be \label{4.7}
\vp_{yy}=c\vp^3, \quad c=\pm 2
\ee
where $c$ is the integration constant which can be reduced to 2
(for $c > 0$) or to -2 (for $c<0$) by scaling the dependent
variable $\vp$. We make such scaling to simplify the following
formulae.

In accordance with (\ref{2.2}), (\ref{4.5}), any solution $\vp$ of
(\ref{4.7}) generates a solution for (\ref{4.2}) of the following
form
\be \label{4.8}
u=\frac{2x \vp_y}{\vp}.
\ee

An explicit solution of equation (\ref{4.7}) for $c=2$ is the
Jacobi elliptic function
 \be\label{4.81}\vp(y)={\rm ds} \lo y, \frac{1}{\sqrt{2}}\ro,\
y=x^2+6t\ee
so (\ref{4.8}) can be represented as
 \be \label{4.9} u=u_1=\frac{2x {\rm cs}\left(y,\frac{1}{\sqrt
2}\right)}{{\rm dn}\lo y,
\frac{1}{\sqrt{2}}\ro}. \ee

The case $c=-2$ leads to the same solution as given in (\ref{4.9}).

\begin{figure}[th]
\centerline{\includegraphics{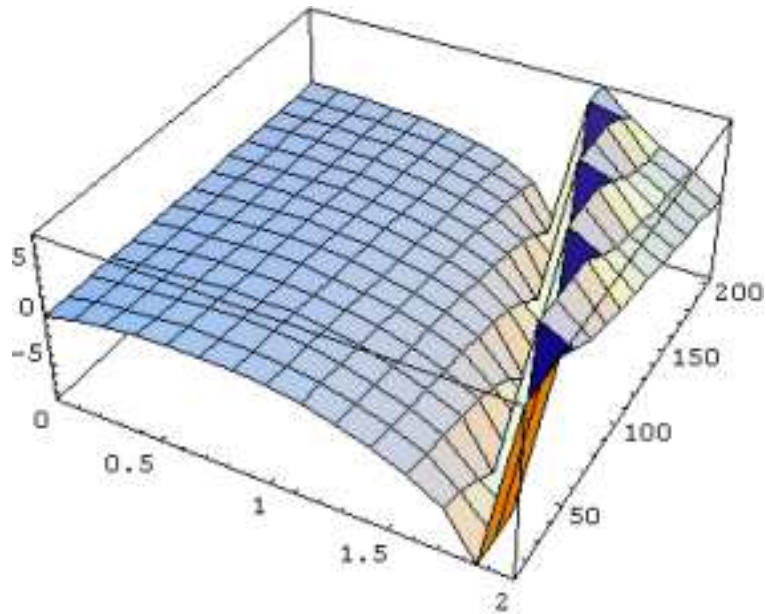}}
\caption{Solution (\ref{4.9}) for equation (\ref{4.2}),\ \ $0< t\leq 200$}\label{f2}
\end{figure}

The solution (\ref{4.9}) is missing in the table of elliptic
function solutions
present in \cite{Clarkson}. The plot of this solution is given by Fig. \ref{f2}.

To construct more elliptic function solutions for (\ref{4.2}) we
notice
that (\ref{4.3}) can be rewritten as
\be\label{4.100}
u=y_x \vp(y)
\ee
with the same functions $\vp$ and $y$ as given in (\ref{4.81}).
Moreover, {\it solution
(\ref{4.9}) can be obtained from  (\ref{4.100}) by the change}
$\vp\to\frac{\vp_y }{\vp}$. And it is
 this observation which opens the way to construct an infinite
number of exact solutions for
(\ref{4.2}). To do this we will exploit some properties of the
elliptic functions formulated
in the following assertions.

{\bf Proposition 1.} {\it Let $\vp=\vp^{(n)}$ be a solution of
equation
(\ref{4.7}) for $c=2$ or $c=-2$. Then \be \label{4.10}
\vp^{(n+1)}=\frac{\vp^{(n)}_y}{\vp^{(n)}} \ee also satisfies this
equation for $c=2$.}

Proof of this and the following propositions is reduced to a direct
verification. We notice that equation
(\ref{4.7}) is equivalent to the following one
\be \label{4.11}
\lo\vp^{(n)}_y\ro^2=\lo\vp^{(n)}\ro^4+C_{n}
\ee
where $C_n$ is the integration constant. Then $\vp^{(n+1)}$ of
(\ref{4.10})
satisfies the equation
\[
(\vp^{(n+1)}_y)^2=(\vp^{(n+1)}_y)^4+C_{n+1}, \ \ \  C_{n+1}=-4C_n.
\]

{\bf Proposition 2.} {\it Let $\vp^{(n)}$ be a solution of equation
(\ref{4.11}) for $C_n>0$. Then this equation is solved also by
the following function}
\[ {\tilde\vp}^{(n)}=\frac{\sqrt{C_n}}{\vp^{(n)}}.\]

{\bf Proposition 3.} {\it Let
$\vp^{(n)}$ be a solution of equation (\ref{4.11}) for
$C_n=-B_n<0$. Then the function \be {\hat
\vp}^{(n)}=\frac{\sqrt{B_n}}{\vp^{(n)}} \label{4.13}\ee satisfies equation
(\ref{4.7}) for $c=-2$ and the following relation \be\label{4.14}
{\lo{\hat \vp}^{(n)}_y\ro}^2=-{\lo{\hat \vp}^{(n)}\ro}^4+B^2_n. \ee
}

Using Propositions 1 and 2 and starting with (\ref{4.100}) we
obtain
infinite
 sets of solutions for equation (\ref{4.2}):
\be\label{4.15} u_n=2x\vp^{(n)}, \ \ n=0,1,2, \cdots, \ee and
\be\label{4.16} {\tilde u}_{2k+1}=\frac{2^{k+1} x}{{\tilde
\vp}^{(2k+1)}}, \ \ k=0,1, 2, \cdots\ee where
$\tilde\vp^{(2k+1)}$ and $\vp^{(n)}$ are defined by (\ref{4.13}) and
the following recurrence relations
\be\label{4.17} \vp^{(n)}=\frac{\vp_y^{(n-1)}}{ \vp ^{(n-1)}} ,  \
\ \vp^{(0)}={\rm ds}\lo y,{\ds \frac {1} {\sqrt 2}} \ro. \ee For
$n=0,1$ we obtain from (\ref{4.15}), (\ref{4.17}) the solutions
given by (\ref{4.3}), (\ref{4.81}), while for $n=2,3, \cdots$ and
$k=0,1, \cdots$ we have
 \be \label{4.18} \ba{l} u_2={\ds
2x\left[\frac{{\rm cd}\lo y,\frac{1}{\sqrt{2}}\ro-{\rm dc}\lo y,
\frac{1}{\sqrt{2}}\ro} {{\rm sn}\lo y, \frac{1}{\sqrt{2}}\ro}-{\rm
cn}\lo y,\frac{1}{\sqrt{2}}\ro{\rm ds}\lo y,
\frac{1}{\sqrt{2}}\ro \right]},\\
\
\\
{\ds u_3=\frac{2x \lo {\rm cs}^4\lo y, \frac{1}{\sqrt{2}}\ro-{\rm
dn}^4\lo y, \frac{1}{\sqrt{2}}\ro\ro}{{\rm dn}\lo y,
\frac{1}{\sqrt{2}}\ro {\rm cs}\lo y, \frac{1}{\sqrt{2}}\ro\lo
\frac{9}{4}\sqrt{2} {\rm cn}^2\lo y, \frac{1}{\sqrt{2}}\ro-{\rm
ds}^2\lo y, \frac{1}{\sqrt{2}}\ro\ro}},
 \\
\cdots, \\
\ea\ee
and
 \be\ba{l} \label{4.19}
\ds {\tilde u}_1= \frac {2x {\rm dn}\left(y, \frac
{1}{\sqrt{2}}\right)} {{\rm cs}\left(y,
\frac{1}{\sqrt{2}}\right)}, \\
\
\\
{\tilde u}_3=\ds {\frac{4x{{\rm dn}\lo y, \frac{1}{\sqrt{2}}\ro
{\rm cs}\lo y, \frac{1}{\sqrt{2}}\ro\lo \frac{9}{4}\sqrt{2} {\rm
cn}^2\lo y, \frac{1}{\sqrt{2}}\ro-{\rm ds}^2\lo y,
\frac{1}{\sqrt{2}}\ro\ro}}{ {\rm cs}^4\lo y,
\frac{1}{\sqrt{2}}\ro-{\rm
dn}^4\lo y, \frac{1}{\sqrt{2}}\ro }},\\
\cdots . \ea \ee
Formulae (\ref{4.10}), (\ref{4.18}) , (\ref{4.19}) and the
recurrence relations (\ref{4.17}) add the list of elliptic
function solutions for equation (\ref{4.2}), found in
\cite{Fush2}, \cite{Clarkson}. Moreover, taking into account the
transparent invariance of (\ref{4.2}) with respect to
displacements of independent variables $t$ and $x$ we can write
more general solutions changing $x\to x +k_1$, $t
\to t + k_2$ with arbitrary constants $k_1$ and $k_2$.

\begin{figure}[th]
\centerline{\includegraphics{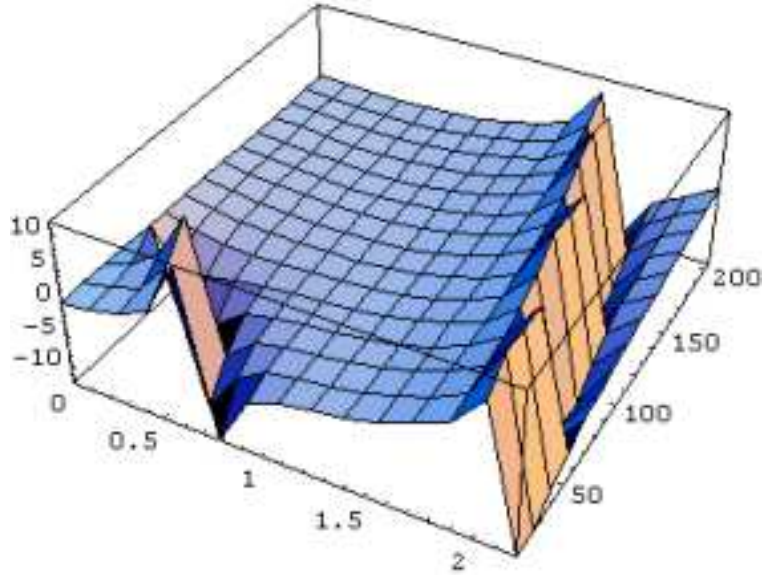}}
\caption{Solution $\tilde u_1$ (\ref{4.19}) for equation (\ref{4.2})}\label{f3}
\end{figure}

The plots of solution $\tilde u_1$ is given in Fig. \ref{f3}.

Propositions 1 and 2 make it possible to construct infinite sets of
exact
 solutions for other equations of the type (\ref{2.4}).

Consider first equations (\ref{2.4}) for $n=3,
\lambda_2=\lambda_3=\lambda_4=0$, i.e.,
\be\label{4.20}
u_t-u_{xx}=-2\lo u^3+\lambda_1 u\ro
\ee
where without loss of generality we can set $\lambda_1=\pm 1$.
For $\lambda_1=-1$ (\ref{4.20}) is equivalent to the Newell-
Whitehead \cite{Newell} equation up to scaling variables $t$ and
$x$.
The Ans\"atze
\be\ba{l}\label{4.21}u=\xi_x\vp(\xi), \quad \xi=k_1
\cosh (x+k_2) \exp(3t), \quad \lambda_1=1, \ea\ee \be\label{4.210}
u=\eta_x\vp(\eta), \quad \eta=k_1 \cos (x+k_2) \exp(3t ), \quad
\lambda_1=-1 \ee reduces (\ref{4.20}) to the form (\ref{4.7}). Thus
repeating the arguments which follow equation (\ref{4.7}) we
come to exact solutions for (\ref{4.20}). The explicit form of
these solutions can be obtained from
(\ref{4.3}), (\ref{4.15})-(\ref{4.19}) via
the changes $y\to \xi, \ 2x \to \xi_x$  for $\lambda_1 =1$ and
$y\to\eta, \ 2x\to \eta_x$ for $\lambda_1=-1$.

Finally, we notice that Proposition 3 makes it possible to
construct
infinite sets of exact solutions for the equations (\ref{2.4}) with
$n=\lambda_4=-1,\ \lambda_2=\lambda_3=0$, i.e., for the equations
 \be\label{4.22}
u_t-u_{xx}=2 u^3 \ee and \be\label{4.23} u_t-u_{xx}=2\lo
u^3+\var u\ro \ee which differ from (\ref{4.2}) and (\ref{4.20})
by the sign of the l.h.s. terms. Indeed, Ans\"atze (\ref{4.100})
and (\ref{4.21}) or (\ref{4.210}) reduce the corresponding
equations (\ref{4.22})
and (\ref{4.23}) to the following equation for $\vp$
\be\label{4.24}\vp''=-2\vp^3\ee where the double prime denotes the
second derivative w.r.t. the corresponding variable (i.e., $y,\
\xi$ or $\eta$).

In accordance with Proposition 3, exact solutions for
(\ref{4.24}) have
the form (\ref{4.13}) where
$\vp^{(n)} $ are solutions of equations  (\ref{4.17}) for {\it
even} $n$. The related list of exact
solutions for equation (\ref{4.22}) is given by the following
formulae:
\be\ba{l}\label{4.25}
{\hat u}_{0}=x{\rm sd}\lo y, \frac{1}{ {\sqrt 2}}\ro,\\
{\hat u}_{2}={\ds \frac {4x{{\rm sn}\lo y, \frac{1}{ {\sqrt
2}}\ro}}{{{\rm cd}\lo y, \frac{1}{ {\sqrt 2}}\ro -{\rm dc}\lo y,
\frac{1}{ {\sqrt 2}}\ro -{\rm cn}\lo y, \frac{1}{ {\sqrt
2}}\ro{\rm dn}\lo y, \frac{1} { {\sqrt 2}}\ro{\rm sn}\lo y,
\frac{1}{ {\sqrt
2}}\ro}}},\\
\cdots\\
{\hat u}_{2k}=\frac{2^kx}{ \vp^{(2k)}} \ea\ee where $\vp^{(2k)}$
are defined by recurrence relations  (\ref{4.17}).

\begin{figure}[th]
\centerline{\includegraphics{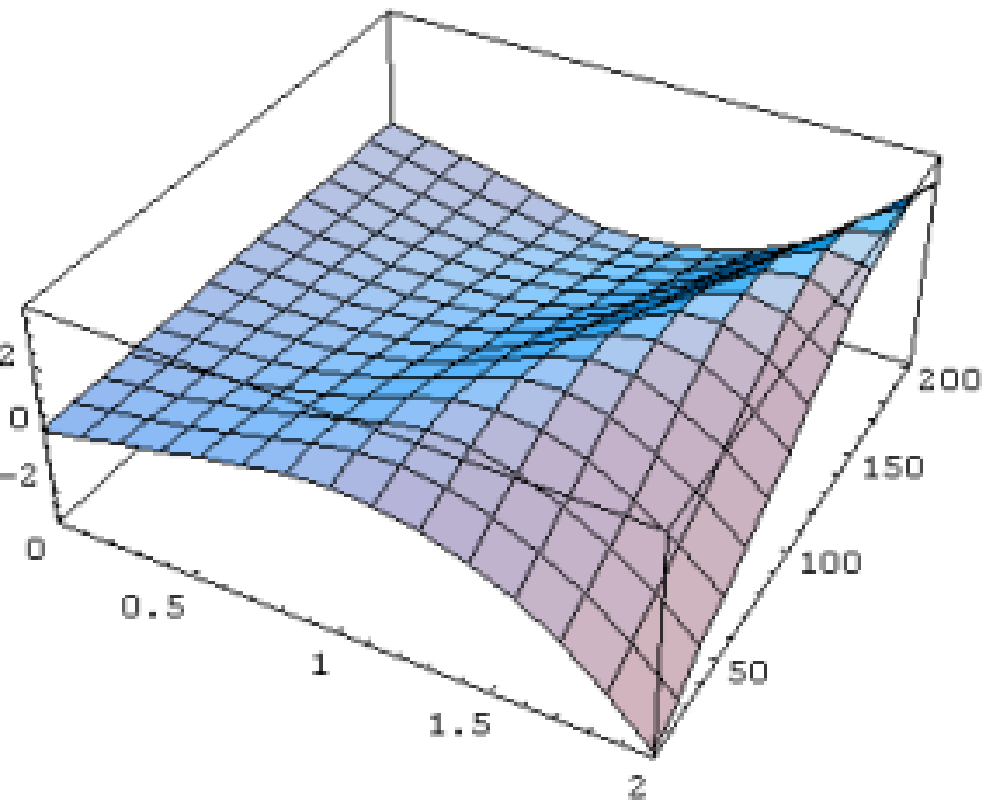}}
\caption{Solution $\tilde u_0$ (\ref{4.19}) for equation (\ref{4.2})}\label{f4}
\end{figure}

\begin{figure}[th]
\centerline{\includegraphics{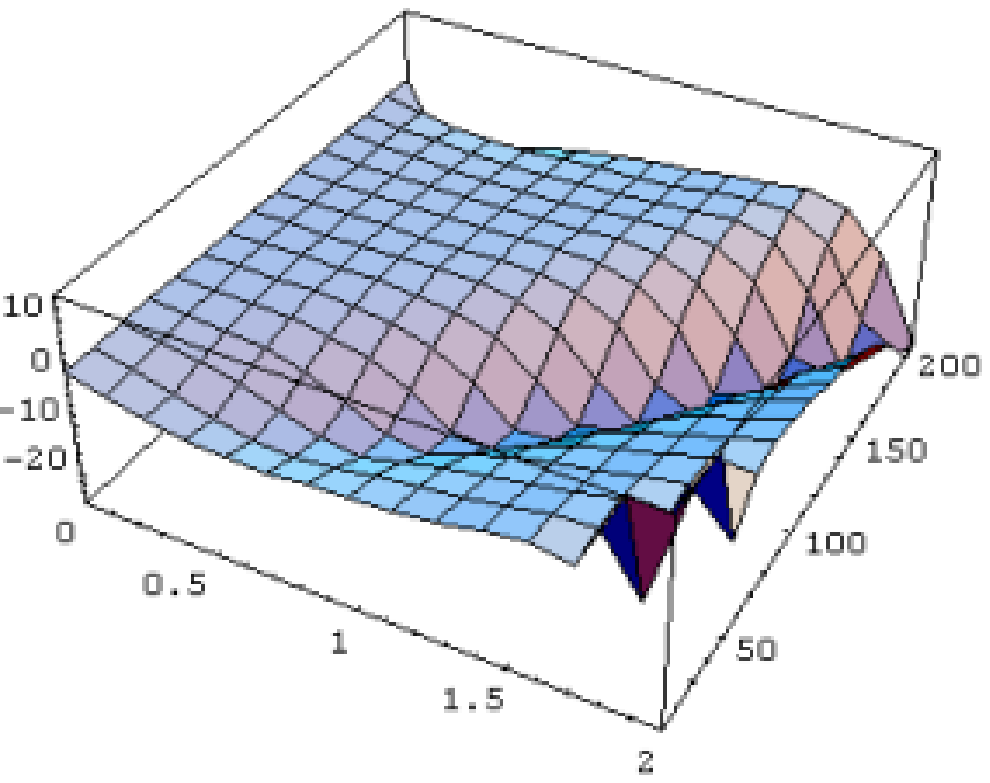}}
\caption{Solution $\tilde u_2$ (\ref{4.19}) for equation (\ref{4.2})}\label{f5}
\end{figure}

The plots of solutions (\ref{4.24}) are given in Figs. \ref{f4}, \ref{f5}.


Solutions
for  (\ref{4.23}) can be obtained from  (\ref{4.25}) by changing
$y\to \xi, \ x\to \xi_x$ and $y\to \eta, \ x\to \eta_x$ for
$\var=1$ and $\var=-1$ respectively.

\renewcommand{\theequation}{\arabic{section}.\arabic{equation}}
\setcounter{equation}{0}

\section{Solutions for arbitrary $n$}

Let us consider equation (\ref{2.4}) with arbitrary $n$ and
construct its exact solutions. In this section  we restrict
ourselves to the case $\lbd_3=\lbd_4=0$ and use a
reduced version of the related potential equation
 (\ref{2.6})
i.e.,
\be \label{5.1} (n-1)  z_t= (n+3)
z_{xx}-(n-1)(\lbd_1 z+ \lbd_2 z_x), \ee \be \label{5.2} 4 z_x
z_{xxx}+(n-3) z^2_{xx}-(n-1)(\lbd_1 z^2_x+\lbd_2 z_x z_{xx}). \ee

Any solution of the system (\ref{5.2}) satisfies (\ref{2.6}) with
$\lbd_3=\lbd_4=0$, the inverse is not true.

To solve (\ref{5.1}), (\ref{5.2}) we
introduce the new variable $y=\frac{z_{xx}}{z_x}$. Then, dividing
(\ref{5.2}) by $z^2_x$ and using the identity
$\frac{z_{xxx}}{z_x}=y_x+y^2$ we transform this equation to the
Riccatti form \be \label{5.3}
y_x+\frac{n+1}{4}y^2-\frac{(n-1)}{4}(\lbd_2 y+\lbd_1)=0 .\ee

Differentiating $y$ w.r.t. $t$ and using (\ref{5.1}) and
(\ref{5.3}) we obtain the following differential consequence
\be\label{5.4} \dot y=A y^3+B y^2+Cy+D \ee where
\[
A=\frac{1}{8}\frac{(n+3)(n-3)(n+1)}{n-1}, \quad
B=\lbd_2\left(1-\frac{3}{16}(n^2-1)\right),
\]
\[
C=\frac{1}{16}\left((n-1)^2\lbd^2_2-2(n+3)(n-3)\lbd_1\right), \quad
D=\frac{(n-1)^2}{16}\lbd_1 \lbd_2.
\]

For arbitrary $n$ the system of equations (\ref{5.3}), (\ref{5.4})
is compatible  but has constant solutions only. In three
exceptional cases $n=\pm3$ and $n=-1$ the compatibility conditions
for (\ref{5.3}), (\ref{5.4}) are less restrictive in as much as the related
coefficient $A$ in (\ref{5.4}) is equal to zero.

Let $n\not=\pm 3$ and $n\not=-1$, then $y=c_1=\mbox{const}$, and
equations (\ref{5.3}), (\ref{5.4}), reduce to the only condition
\[ \lbd_1=-c_1\lbd_2+(k+1)c_1^2, \quad
k=\frac{2}{n-1}. \]
The corresponding solution for the system
(\ref{5.1}), (\ref{5.2}) is \be \label{5.6} z=e^{c_1x+k c_1^2
t}+c_2 e^{\lo\lbd_2 c_1-(k+1)c_1^2\ro t} \ee and the related exact
solution (\ref{2.2}) takes the form
\be \label{5.7}
u=\frac{c_1^{k}}{\lo 1+c_2e^{-c_1x-\lo(2k+1)c_1^2-\lbd_2c_1\ro
t}\ro^{k}}.
\ee

Thus we find exact solutions (\ref{5.7}) for the equation \be
\label{5.60}  u_t-u_{xx}=-k(k+1)u^n+\lbd_2k
u^\frac{n+1}{2}+((k+1) c_1^2-\lbd_2c_1){k u}. \ee
This equation
belongs to the Kolmogorov-Petrovski-Piskunov type provided \be
\label{5.9} \lbd_2=(k+1)(c_1+1). \ee The corresponding plane wave
solution (\ref{5.7}) propagates with the velocity
\[v=k+1-k c_1. \]

In special cases $n=\pm 3$ and $n=-1$ equations (\ref{5.1}),
(\ref{5.2}) admit more extended classes of exact solutions.
 In particular, for $n=3$ we can recover exact
solutions for (\ref{1.3}) caused by conditional symmetry and
classical Lie symmetry as well. We will not study these special
cases here.
\renewcommand{\theequation}{\arabic{section}.\arabic{equation}}
\setcounter{equation}{0}

\section{ Solitary wave solutions}

Consider now the general equation (\ref{2.3}) with arbitrary
parameters $\lbd_1$, $\lbd_2$, $\lbd_3$ and $\lbd_4$. It seems to
be impossible integrate in closed form the related potential
equations (\ref{2.6}).
 Here we search for particular solutions which
belong to soliton type and so have good perspectives for various
applications.

Let us consider solutions for (\ref{2.3}) of the form
$z(t,x)=U(\xi)$
where  $\xi=\mu  t+x$ and $
\mu$ is an arbitrary (nonzero) constant. Then we come to the
following ordinary differential equation for $U$
\be\label{6.0}\ba{l}
U[U'(\mu U''-U'''-\lbd_3 U)-\lbd_4 U^2-(k-1)(U'')^2]\\
=(U')^2[(\mu+\lbd_2)U'+\lbd_1 U-(2k+1)U'']\ea\ee
where $U'=\frac{dU}{d\xi}$.

Let us follow \cite{fan} and search for solutions for
(\ref{6.0}) in the form
\be\label{6.01}U=\nu_0+\nu_1\vp+\nu_2\vp^2+\cdots,\ee
where $\nu_0, \ \nu_1, \cdots$ are constants and $\vp$ satisfies
equation of the following general form
\be\label{6.02} \vp'=\ve {\sqrt {c_0+c_1\vp+c_2\vp^2+\cdots}}
\ee
where $\ve=\pm 1$. In order (\ref{6.01}) be compatible with
(\ref{6.0}) we have to equate separately the terms which include
odd and even powers of
the square root given by (\ref{6.02}). In view of this
 we come to the following system
\be\label{6.03}
U'(\mu UU''-\lbd_3 U^2)=(U')^3(\mu+\lbd_2),\ee
\be\label{6.04}
U(U'U'''+\lbd_4 U^2+(k-1)(U'')^2)=(U')^2((2k+1)U''-\lbd_1 U).\ee

Dividing any term in (\ref{6.03}) by $\mu U^2U'$ we come to the
Riccatti
equation
\[ Y'-\frac{\lbd_2}{\mu}Y^2=\frac{\lbd_3}{\mu}\]
for $Y=\frac{U'}{U}$, whose general solutions are
\be\label{6.05} \ba{l}Y=
{\sqrt{\frac{-\lbd_3}{\lbd_2}}}\tanh\lo\frac{\sqrt{-\lbd_2\lbd_3}}
{\mu}\xi+C\ro, \\
Y={\sqrt{\frac{-\lbd_3}{\lbd_2}}}\lo\tanh\lo\frac{\sqrt{-\lbd_2\lbd_3}}
{\mu}\xi+C\ro\ro^{-1}, \ \  {\rm if}\ \  \lbd_2\lbd_3<0,\ea\ee
\be\label{6.005}
y={\sqrt{\frac{\lbd_3}{\lbd_2}}}\tan\lo\frac{\sqrt{\lbd_2\lbd_3}}
{\mu}\xi+C\ro, \ \ \ \ \ {\rm if}\ \  \lbd_2\lbd_3>0,\ee
\be\label{6.0005}
y=-\frac{\mu}{\lbd_2( \xi+C)}, \ \ \ \ \ \ \ {\rm if}\ \
\lbd_3=0\ee
where $C$ is the integration constant.

Thus all solutions for (\ref{6.0}) which can be obtained with using
the algebraic method \cite{fan} are exhausted by hyperbolic,
triangular and rational ones presented by relations (\ref{6.05})-
(\ref{6.0005}).

Solutions (\ref{6.05}), (\ref{6.005}) and (\ref{6.0005}) are
compatible with
(\ref{6.04}) provided
\[ \mu=-\lbd_2,\ \lbd_1=-k\frac{\lbd_3}{\lbd_2}, \
\lbd_4=(1-k)\lo\frac{\lbd_3}{\lbd_2}\ro^2, \ \lbd_2\lbd_3\neq 0\]
and
\[\lbd_1=\lbd_4=0,\ \lbd_3=0\]
respectively. Using variables
\[
\tau=\frac{2}{(n-1)^2}t,\ y=\frac{{\sqrt 2}}{n-1}x,\
\sigma=-\lbd_2(n-1),\
\nu=\frac{\lbd_3}{\lbd_2}\]
we can rewrite the related equation (\ref{2.4}) as follows:
\be\label{6.5}
u_\tau-u_{yy}=\lo 1+\nu u^{1-n}\ro\lo-(n+1)u^n+\nu(n-3)u+
\sigma u^\frac{n+1}{2}\ro.\ee
The corresponding solutions (\ref{2.2}),  (\ref{6.05}) for equation
(\ref{6.5}) have the following form
\be \label{6.40}\ba{l}
u=(-\nu)^{\frac{1}{n-1}} \lo\tanh
\left(b\left(y-\frac{\sigma}{\sqrt 2}
t\right)+C\right)\ro^\frac{2}{n-1},\ea\ee
\be\label{6.400}\ba{l}u=(-\nu)^{\frac{1}{n-1}} \lo\tanh
\left(b\left(y-\frac{\sigma}{\sqrt 2}
t\right)+C\right)\ro^\frac{2}{1-n}\ea\ee
where $\nu<0$ and $b=(n-1){\sqrt\frac{-\nu}{2}}$,
\be\label{n1}
u=(\nu)^{\frac{1}{n-1}} \lo\tan
\left(b\left(y-\frac{\sigma}{\sqrt 2}
t\right)+C\right)\ro^\frac{2}{n-1}\ee
where $\nu>0$ and  $b=(n-1){\sqrt\frac{\nu}{2}}$, and
\be\label{n2}u=2^{n-1}\lo(n-1)\lo y-
\frac{\sigma}{\sqrt 2}\tau+C\ro\ro^\frac{2}{1-n}\ee
if $\nu=0$.

 For $\frac{2}{n-2}>1$ formula (\ref{6.40})
presents nice solitary wave solutions which propagate with the
velocity $\frac{\sigma}{\sqrt 2}$. In the case $n=2$ we come to the
bell-shaped
solitary wave solution which will be discussed in Section 7.

If $\frac{2}{n-2}<-1$ then (\ref{6.40}) is a singular solution whose
physical relevance is doubtful. However, in this case
equation (\ref{6.5}) admits another solitary wave
solutions which are given now by relation (\ref{6.400}).

We see  such solutions exist for the
extended class of the nonlinear reaction-diffusion equations
defined by formula (\ref{6.5}).


\renewcommand{\theequation}{\arabic{section}.\arabic{equation}}
\setcounter{equation}{0}

\section{Exact solutions for the Fisher equation}

Let us return to Section 4 and consider in more detail the
important case $n=2$. Setting in (\ref{5.60}) $\lbd_2=0, \ c_1=-1$
and making
the change
\be\label{7.0}\tau=6t,\ \  y=\sqrt{6} x \ee we come to the Fisher
equation
(\ref{1.1}) for $u(\tau,y)$:
\be\label{7.1}
u_{\tau}-u_{yy}=u(1-u).\ee

Thus the Fisher equation is a particular case of (\ref{5.60}) and
so
our solutions (\ref{5.7}) are valid for (\ref{7.1})
provided we make the above mentioned changes of variables and set
$c_1=-1$ in
accordance with (\ref{5.9}). As a result we recover the well-known
Ablowitz-Zeppetella \cite{ab} solution
\be \label{7.3}
u=\frac{1}{\lo 1+ c_2 e^{
\frac{y}{\sqrt{6}}-\frac{5\tau}{6}}\ro^2}. \ee
This
solution can be expressed via hyperbolic functions
\be\label{7.4}  u= u_1=\frac{1}{4}\left(1-\tanh
\left(\frac{y}{2\sqrt{6}}- \frac{5}{12}\tau-c\right)\right)^2,
\quad c=\frac{1}{2} \ln |c_2|, \ee
\be\label{7.5} u=
 u_2=\frac{1}{4}\left(1-\coth \left(\frac{y}{2\sqrt{6}}-
\frac{5}{12}\tau-c\right)\right)^2 \ee for $c_2>0$ and $c_2<0$
respectively.

Taking into account the symmetry of (\ref{7.1}) w.r.t. the discrete
transformation $ u \to 1 -  u$ we obtain two more
solutions: $u_3=1-u_1 $ and $ u_4=1-u_2$.
Finally, bearing in mind the symmetry of (\ref{7.1}) w.r.t. the
space
reflection $y \to -y$ we come to four more exact solutions by
changing $y \to -y$ in $u_1, u_2, u_3$ and $u_4$.

Thus starting with our general formulae (\ref{5.6}) and (\ref{5.7})
we
come to the family of eight exact solutions for the Fisher
equation.
All of them are plane
waves propagating with the velocity $\pm\frac{5}{\sqrt{6}}$.

To find additional exact solutions we use the Ans\"atz
(compare with (\ref{4.100}))
\be { u}= 3{z_y^2}\vp(z)\label{7.10}\ee
where $\vp$ and $z$ are functions to be found.
Substituting (\ref{7.10}) into (\ref{7.1}) we come to the following
reduced equations
\be\ba{l} z_\tau=5z_{yy},\\
4z_yz_{yyy}-z_{yy}^2=\frac12 z_y^2\ea\label{7.11}\ee
and
\be\label{7.12}\vp_{zz}=3\vp^2.\ee

We see that $\vp$ has to satisfy the Weierstrass equation
(\ref{7.12})
which we rewrite in the following equivalent form
\be  {\tilde \vp}_z^2=4{\tilde\vp}^3-C,\ {\tilde\vp}=\frac12 \vp
\label{7.13}\ee
where $C$ is the integration constant.

Starting with (\ref{7.10}) and choosing the following exact
solutions of (\ref{7.11}) and (\ref{7.13}):
\be z=\exp\lo-\frac{1}{\sqrt 6}y+\frac56 \tau\ro, \
\vp=\frac{2}{z+k}\label{7.14}\ee
we come to the Ablowitz-Zeppetella solutions (\ref{7.3}) for the
Fisher
equation.

We notice that relations (\ref{7.14}) present only a very
particular solution of
(\ref{7.13}) which correspond to zero value of the parameter $C$.
 In addition, there exist the infinite number of other solutions
corresponding
to non-zero $C$. The related functions (\ref{7.10}) are:
\be {u} =\frac12 z^2\wp(z, 0, C), \ \
z=\exp\lo-\frac{1}{\sqrt 6}y+\frac56 \tau+k\ro\label{7.15}\ee
where $\wp(z, 0, C)$ is the Weierstrass function satisfying
equation
 (\ref{7.13}) for $C\neq 0$.

\begin{figure}[th]
\centerline{\includegraphics{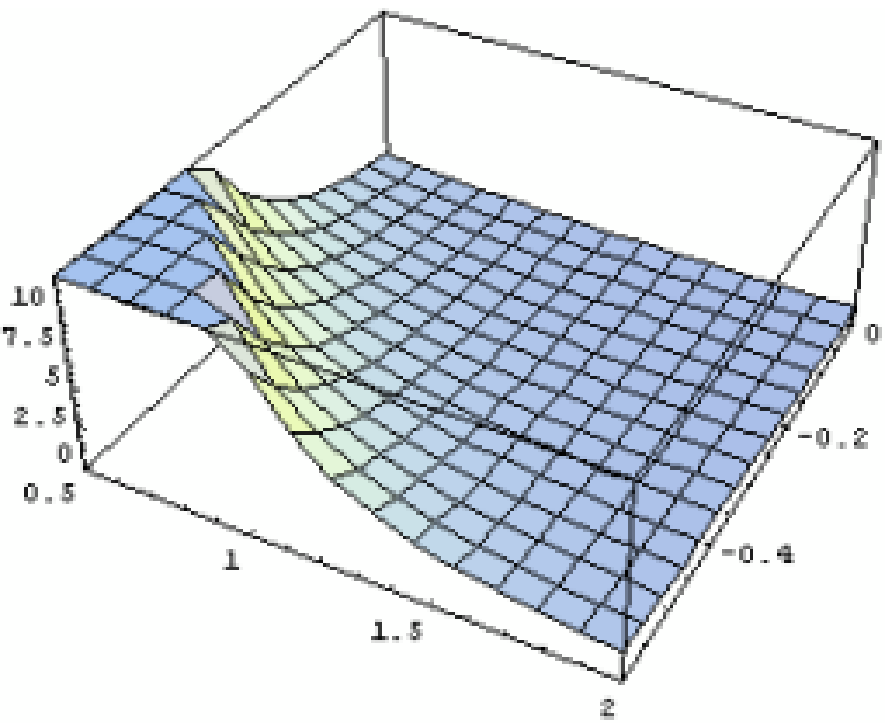}}
\caption{Solution (\ref{7.15}) with $k=0, C=10^2$ for the Fisher equation (\ref{7.1})}\label{fish1}
\end{figure}

\begin{figure}[th]
\centerline{\includegraphics{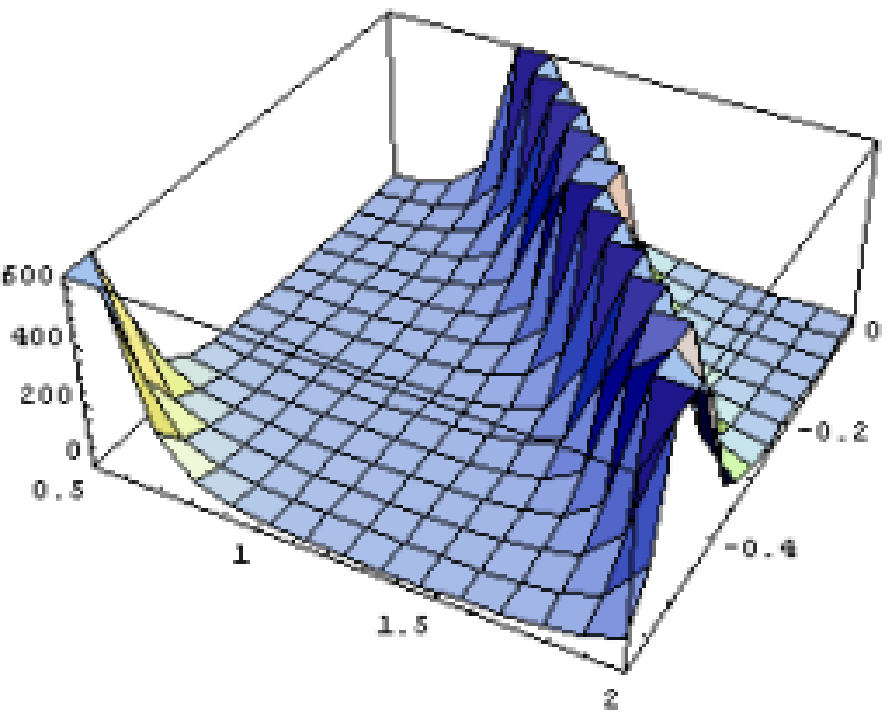}}
\caption{Solution (\ref{7.15}) with $k=0, C=10^4$ for the Fisher equation (\ref{7.1})}\label{fish2}
\end{figure}

\begin{figure}[th]
\centerline{\includegraphics[width=10.16cm,height=8.22cm]{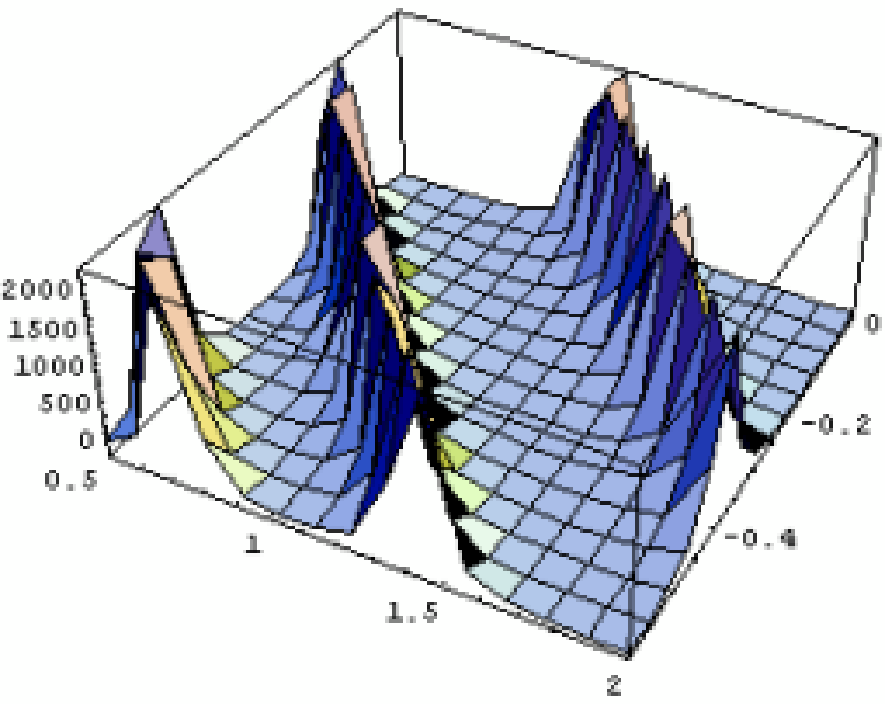}}
\caption{Solution (\ref{7.15}) with $k=0, C=10^6$ for the Fisher equation
(\ref{7.1})}\label{fish3}
\end{figure}

In order to solutions (\ref{7.10}) be bounded it is sufficient to restrict ourselves
to the case when $-\frac{1}{\sqrt 6}y+\frac56 \tau+k>0$. Such conditions can be satisfied,
e.g., for arbitrary positive $y$ and negative $\tau$ and $k$. The related solutions can be
interpreted as ones describing the history of the process because the time variable takes
arbitrary negative values. The graphics of solutions (\ref{7.10}) for some values of the parameter
$C$ are given by Figures \ref{fish1}-\ref{fish3}.

Thus the Fisher equation admits the infinite set of exact
 solutions which include the Ablowitz-Zeppetella solutions
 (\ref{7.3}) and also solutions (\ref{7.15}) enumerated by two
parameters, $C\neq 0$ and $k$.  All these solutions are
plane waves propagating with the same velocity $v=\frac{5}{\sqrt
6}$.

In the following section we consider generalized Fisher equations
which admit exact solutions with arbitrary propagation velocities.

\renewcommand{\theequation}{\arabic{section}.\arabic{equation}}
\setcounter{equation}{0}

\section{Generalizations of the Fisher equations}

Let us consider equation (\ref{6.5}) for $n=2$, which takes the
following form
\be \label{6.6}
u_\tau-u_{yy}=(u+\nu)(-3u+\sigma\sqrt{u}-\nu).
\ee

Relation (\ref{6.6}) is a formal generalization of the Fisher
equation in as much as in the case $\sigma=0$ (\ref{6.6}) is
equivalent to
(\ref{1.1}). However, for $\sigma\neq 0$ equation (\ref{6.6})
admits
soliton solutions (\ref{6.40}) and (or) solutions
(\ref{6.400})-(\ref{n2}) and so has absolutely another nature
then (\ref{1.1}). Nevertheless for small $\sigma$ it would be
interesting to treat (\ref{6.6}) as a perturbed version of
(\ref{1.1}).

Consider equation (\ref{6.6}) in more detail. Let $\nu<0$ then
scaling dependent and independent variables we can reduce its value
to the following one
\be\label{8.2}
\nu\to\nu'=-\frac32, \ {\rm if} \ \nu<0.\ee
Setting then
\be\label{8.3}
\tilde u=\frac 32 -u, \ \sigma=3\ve, \ t=3\tau, \ x=-{\sqrt 3}y\ee
we come to the following relation
\be\label{8.4}
\tilde u_t-\tilde u_{xx}=\tilde u\lo 1-\tilde u +\ve\lo\frac 32 -
\tilde u\ro^\frac 12\ro.\ee

In the limiting case $\ve\to 0$ equation (\ref{8.4}) reduces to the
Fisher equation in the canonical formulation (\ref{1.1}).

In accordance with Section 5 equation (\ref{6.6}) admits  nice
bell-shaped traveling wave solution
(\ref{6.40}) which transforms via changes (\ref{8.2}), (\ref{8.3})
to the following form
\be\label{8.5}
\tilde u=\frac{3}{2\cosh^2\lo\frac 12\lo x-\frac{\ve}{\sqrt
6}t\ro+C\ro}.\ee

Consider now equation (\ref{6.6}) for $\nu\sigma=0=0$ and set
$u=\frac{\tilde u}{3}$. As a result we reduce
(\ref{6.6}) to the simplest form
\be\label{8.7}
\tilde u_\tau-\tilde u_{yy}=-{\tilde u}^2.\ee
 The Ans\"atz
\[
u=\frac{z_x^{2}}{6z^2}+\frac
13\left(1+\var\sqrt{\frac{3}{2}}\right)\frac{z_{xx}}{z},
\quad \var=\pm 1
\]
leads to the following reduced equations
\[
z_{xxx}=0, \quad z_t=\kappa z_{xx}, \quad \kappa=5(3\pm \sqrt{6}).
\]
Thus we have $z=\frac{x^2}{2}+\kappa t$ and the related exact
solution
for
(\ref{8.7}) is
\[
u=
\frac{(3\pm \sqrt{6})x^2+10(12\pm 5\sqrt{6})t}{3\left(x^2+
10(3\pm \sqrt{6})t\right)^2}.
\]
We notice that this solution can be found also using the classical
Lie reduction.

Finally, let us consider one more generalization of the Fisher
equation given
by relation (\ref{5.60}) for $n=2$. Using notations (\ref{7.0}) we
rewrite it
 in the following form
\be \label{8.70} \tilde u_{\tau}-\tilde u_{yy}=\tilde
u(-c_1+(c_1+1) \tilde u^{\frac{1}{2}}- \tilde u). \ee
For $c_1=-1$ equation (\ref{8.70}) reduces to the Fisher equation
(\ref{7.1}).

In accordance with the results presented in Section 4 equation
 (\ref{8.70}) admits exact solutions (\ref{5.7}) which in our
notations can be rewritten as
\be \label{8.8}
\ba{l}
\tilde u_1=\frac{1}{4}\left(1+\tanh
\left(\frac{c_1}{2\sqrt{6}}y+\frac{c_1(2c_1-3)}{12}\tau-c\right)\right)^2,\\
\tilde u_2=\frac{1}{4}\left(1+\coth
\left(\frac{c_1}{2\sqrt{6}}y+\frac{c_1(2c_1-3)}{12}\tau-c\right)\right)^2.
\ea
\ee
Two more solutions can be obtained by changing $y \to -y$ in
(\ref{8.8}).

Formulae (\ref{8.8}) present the analogies of solutions
(\ref{7.4}), (\ref{7.5}) for equation (\ref{8.70}).
In contrast with (\ref{7.4}), (\ref{7.5}) these solutions describe
a wave whose propagation velocity
is equal to $\frac{2c_1-3}{\sqrt{6}}$
Thus
changing parameter $c_1$ in (\ref{8.70}) we can obtain solutions
(\ref{8.8}) with any  velocity of propagation given {\it ad hoc}.
In other words
we always can take this velocity in accordance with experimental
data.

Thus equation (\ref{8.70}) reduces to the Fisher equation if the
parameter $c_1$ is equal to $-1$. Moreover, both equations
(\ref{7.1})
and (\ref{8.70})  admit the analogous exact solutions,
(\ref{7.4}), (\ref{7.5}) and (\ref{8.8}), which, however, have
different propagation velocities.

\renewcommand{\theequation}{\arabic{section}.\arabic{equation}}
\setcounter{equation}{0}

\section{ Discussion}

There exist well known regular approaches to search for exact
solutions of nonlinear partial differential equations - the
classical
Lie approach \cite{Lie}, the conditional (non-classical) symmetries
method \cite{Bluman}, \cite{Wint}, \cite{Fush}, \cite{Fush2},
generalized conditional symmetries \cite{Focas}, etc. These
approaches
present effective tools for finding special Ans\"atze which make it
possible to reduce the equation of interest and find its particular
solutions.

However, sometimes it seems that the Ans\"atze by themselves are
more
fundamental than the related symmetries. First, historically,
the most famous Ans\"atze (like the Cole-Hopt one for the Burgers
equation)
 was found without
a scope of a symmetry approach. Secondly, some of Ans\"atze are
effective
in  rather extended classes of problems characterized by absolutely
different symmetries.  In addition, in some cases the direct search
for the Ans\"atz is a more straight-forward and effective procedure
than search for (conditional) symmetries. We remind that the
conditional symmetry approach presupposes
search for solutions of nonlinear determining equations which
in many cases are not simpler than the equation whose
symmetries are investigated \cite{Renat}.

The present paper is based on using special Ans\"atze (\ref{2.2}),
(\ref{4.100}), (\ref{7.10}) which have the following general form
\be\label{9.1}u=z_x^k\vp(z)\ee
where $z$ is an unknown function of independent variables $t,x$ and
$\vp$ is a function of $z$. The Ans\"atze (\ref{9.1}) appear to
be very effective for the extended class of nonlinear
reaction-diffusion equations. In particular, they make it possible
to find new exact solutions for the very well studied heat
equations with cubic and quadratic polynomial non-linearities.
Moreover, such Ana\"atze can be used to reduce wave equations of
another type, e.g., hyperbolic equations. We plane to discuss the
related results elsewhere.

We present the extended list of exact solutions for the Fisher
equation and for the heat equation with the cubic polynomial
non-linearity. In fact we present an infinite number of different
new solutions which are expressed via Weierstrass functions and
Jacobi elliptic  functions. We believe that these solutions will
be useful for applications, in particular, for analysis of the
related boundary value problems.

We propose a generalization of the Fisher equation which
preserves the type of its exact solutions, but predicts
another propagation velocity. This property differs (\ref{8.70})
from numerous  other generalizations of
the Fisher equation refer, e.g.,  to \cite{Lak} and references
cited
therein.

Finally, we find soliton solutions for a number  of nonlinear
equations (\ref{2.4}). To make this we use the algebraic method
\cite{fan} which however was applied not directly to
the equation of interest (\ref{2.4}) but to the potential equation
(\ref{2.6}). By this we extend the class of non-integrable
equations which have soliton solutions to the case
of appropriate quasi-linear heat equations (\ref{KPP}).

We stress that all these results were obtained with essential using
the Ans\"atz (\ref{9.1}). It seems to be an intriguing problem to
find a regular way for searching such "universal" Ans\"atze.


\begin{thebibliography}{99}

\bibitem{Murry} Murray J D Mathematical Biology, Springer, 1991

\bibitem{KPP} Kolmogorov A N, Petrovskii I G and Piskunov N S 1937
Bull. Moscow Univ. S\'er. Int. A {\bf 1} 1

\bibitem{Fit} Fitzhugh R 1961 Biophys. J. {\bf 1} 445\\
Nagumo J. S., Arimoto S and Yoshizawa S 1962 Proc. IRE {\bf 50}
2061

\bibitem{Newell} Newell A. C. and Whitehead J A 1969 J. Fluid Mech.
{\bf 38} 279


\bibitem{Dorodnitsyn}Dorodnitsyn V A 1982 Comp. Meth. Phys. {\bf
22}
115

\bibitem{Lie} Lie S 1883 Transformationgruppen (in 3 Bds) Leipzig

\bibitem{olver} Olver P 1986 Application of Lie groups to
differential equations, Springer, Berlin


\bibitem{Ron}Nikitin A G and Wiltshire R 2000 in: {\it Symmetries
in Nonlinear Mathematical Physics, Proc. of the Third Int. Conf.
, Kiev, July 12-18, 1999, Ed. A.M. Samoilenko} ( Inst.
of Mathematics of Nat. Acad. Sci. of Ukraine, Kiev);\\
 Cherniha R and King J 2000 J. Phys. A {\bf 33} 257;\\
Nikitin A G and Wiltshire R 2001 J. Math. Phys. {\bf 42} 1666

\bibitem{Bluman} Bluman G W and Cole G D 1969 J. Math. Mech. {\bf
18}
1025

\bibitem{Fush} Fushchych W I and Nikitin A G Symmetries of
Maxwell's
equations, Reidel, Dordrecht, 1987;\\
Fushchych W I 1991 Ukr. Math. Zh. {\bf 43} 1456

\bibitem{Wint} Levi D and Winternitz P 1989 J. Phys. A {\bf 22}
2915

\bibitem{Fush2} Fushchich W I and Serov M I 1990 Dokl. Akad. Nauk
Ukr.
SSR, Ser. A {\bf 4} 24

\bibitem{Clarkson} Clarkson P A and Mansfield E L 1993 Physica D
{\bf
70} 250

\bibitem{fan}Fan E 2002 J. Phys. A {\bf 35} 6853

\bibitem{Hirota}Hirota R and Satsuma J 1981 Phys. Lett. A {\bf 85}
407


\bibitem{ab} Ablowitz M J and Zeppetella A 1979  Bull. Math. Biol.
{\bf 41} 835

\bibitem{Focas}Fokas A S and Liu Q M 1994 Theor. Math. Phys. {\bf
99} 371

\bibitem{Renat} Zhdanov R Z and Lahno V I 1998 Physica D {\bf 122}
178

\bibitem{Lak}Needham D J and King A C 2002 Proc. Roy. Soc. (London)

{\bf 458} 1055;\\Bindu P S, Santhivalavan M and Lakshmanan M 2001
J. Phys. A
{\bf 34} L689.

\end{thebibliography}
\end{document}